\documentclass[showpacs,preprintnumbers,amsmath,amssymb,superscriptaddress,
preprint,psfig]{revtex4}
\usepackage{graphicx}
\draft
\begin{document}
\newcommand {\be}{\begin{equation}}
\newcommand {\ee}{\end{equation}}
\newcommand {\bea}{\begin{eqnarray}}
\newcommand {\eea}{\end{eqnarray}}
\newcommand {\nn}{\nonumber}


\title{Triplet superconductivity in the skutterudite $\rm PrOs_4Sb_{12}$
}

\author{Kazumi Maki}
\address{Department of Physics and Astronomy, University of Southern
California, Los Angeles, CA 90089-0484}

\author{Stephan Haas}
\address{Department of Physics and Astronomy, University of Southern
California, Los Angeles, CA 90089-0484}

\author{David Parker}
\address{Department of Physics and Astronomy, University of Southern
California, Los Angeles, CA 90089-0484}

\author{Hyekyung Won}
\address{Department of Physics, Hallym University, Chunchon 200-702,
South Korea
}

\author{K. Izawa}
\address{Institute for Solid State Physics, University of Tokyo,
Kashiwanoha, Kashiwa, Chiba 277-8581, Japan}

\author{Y. Matsuda}
\address{Institute for Solid State Physics, University of Tokyo,
Kashiwanoha, Kashiwa, Chiba 277-8581, Japan}

\date{\today}
\textrm{\textit{}}

\begin{abstract}

There is mounting evidence for triplet superconductivity in the
recently discovered skutterudite compound $\rm PrOs_4Sb_{12}$.
In this work, we propose nodal
order parameters for the A- and B-phases
of this superconductor which are consistent with angle dependent
magnetothermal conductivity measurements and with low-temperature
thermal conductivity data in the range T $\agt$ 150 mK. 
The quasiparticle density of states and
the thermal conductivity $\kappa_{zz}$ are derived within
the quasiclassical approximation.

\end{abstract}
\pacs{}
\maketitle

\noindent{\bf \it 1. Introduction}

The skutterudite compound $\rm PrOs_4Sb_{12}$ is a heavy-fermion
superconductor with a transition temperature of 1.8K.
\cite{bauer,vollmer,kotegawa} Angle dependent magnetothermal
conductivity measurements on this material have revealed 
an interesting multi-phase structure, characterized by energy
gap functions $\rm \Delta ({\bf k})$ with point nodes.\cite{izawa1,won1}
Previously, several unconventional order parameters,
including s+g-wave symmetry, have been proposed to account for this
nodal structure.
\cite{maki1} Most of these    
models that have so far been considered for 
$\rm PrOs_4Sb_{12}$ are not able to describe self-consistently the observed
angle dependence.\cite{models}

Recently, there has been mounting experimental evidence for
triplet superconductivity in this compound. First, from
$\mu$-SR measurements Aoki et al. discovered a remnant magnetization
in the B-phase of this compound, indicating triplet pairing.\cite{aoki}
Second, thermal conductivity measurements along the [0 0 1]
direction at low temperatures (T $\agt$ 150 mK) indicate T-linear and
H-linear behavior, consistent with a triplet order parameter.\cite{izawa2}
Third, the observed angle dependence of $\kappa_{zz}$ with
$\vec{H}$ rotated within the x-z plane indicates triplet superconductivity.
Finally, recently reported NMR data by Tou et al.\cite{tou1}
for the Knight shift in
$\rm PrOs_4Sb_{12}$ also suggest triplet pairing.

The objective of this paper is twofold. First, we propose a new set
of nodal order parameters for the A and B phases of $\rm PrOs_4Sb_{12}$,
which are able to describe the thermal conductivity data observed in Refs.
\cite{izawa1} and \cite{izawa2}. The proposed model can also
describe the isotropic superfluid density in the B-phase reported by
Chia et al.\cite{chia}, albeit with the provision that it has to be
assumed that the nodal points in this experiment are aligned parallel
to the external magnetic field when the sample is field-cooled.
\cite{won2,parker} In order
to interpret the observed $\theta$-dependence of $\kappa_{zz}$
\cite{izawa2} in terms of the present model, the nodes in the B-phase
have to be aligned parallel to [0 0 1].

The second objective of this work
is to study and make predictions for the nodal
excitations in the vortex state of the proposed model. Simple expressions
will be derived for the quasiparticle density of states as well as for the
angle dependent magnetothermal
conductivity. The thermal conductivity obtained from this model describes the
experimental data\cite{izawa1,izawa2} well, 
whereas the previously proposed
order parameters do not.

\section{Nodal superconductivity in $\rm PrOs_4Sb_{12}$}

Following Ref. \cite{maki1}, we consider energy gap functions
$\rm \Delta ({\bf k})$ with point nodes at [0 1 0] and [0 -1 0] in the
B-phase, and with point nodes at [1 0 0], [0 1 0], [-1 0 0], and [0 -1 0]
in the A-phase. Furthermore, in order to consistently describe the
thermal conductivity data of Ref. \cite{izawa2}
one also needs nodes
at [0 0 1] and [0 0 -1]. These constraints suggest 
\bea
\vec{\Delta}_A({\bf \rm k}) = \frac{3}{2} \hat{d} \Delta e^{\pm i
\phi_1 \pm i \phi_2 \pm i \phi_3} \left( 1 - \hat{k}_1^4 -
\hat{k}_2^4 - \hat{k}_3^4 \right), 
\eea 
with $e^{\pm i \phi_1 } =
\hat{k}_2 \pm i \hat{k}_3$ etc. for the A-phase, and with 
\bea
\vec{\Delta}_B({\bf \rm k}) = \hat{d} \Delta e^{\pm i \phi } \left( 1
- \hat{k}_3^4 \right) 
\eea 
for the B-phase. We note that the p+h
order parameter  for the A phase
satisfies the cubic symmetry. Also, for the
B-phase the symmetry axis is rotated parallel to the crystal
c-axis. Hence, the angular part, $f = 1 - \hat{k}_3^4$, of the B-phase
is the same as for the previously considered s+g superconductor,
whereas the angular part of the A-phase, $f = \frac{3}{2} \left(1
- \hat{k}_1^4 - \hat{k}_2^4 - \hat{k}_3^4\right) $, is different.
The magnitude of these superconducting order parameters 
are shown in Fig. 1. Since both of these triplet order
parameters break chiral symmetry, the ground state of the A-phase
is sixfold degenerate and the ground state of the B-phase is
twofold degenerate.

\begin{figure}[h]
\includegraphics[width=4.0cm]{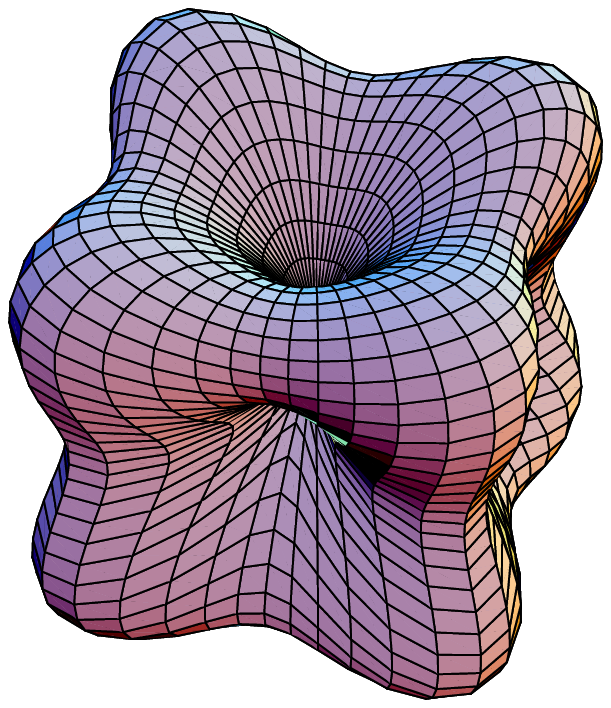}
\includegraphics[width=5.0cm]{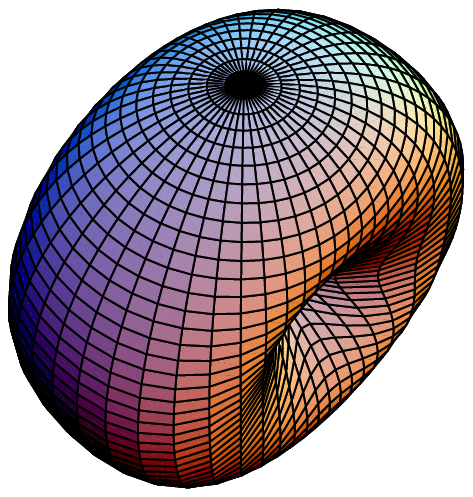}
\vspace{0.3cm}
\caption{
$|\Delta ({\bf k})|$
of the proposed p+h-wave superconducting order 
parameter in the A-phase (left) and in the B-phase (right) 
of $\rm PrOs_4Sb_{12}$.
}
\end{figure}

\section{Quasiparticle spectrum}

In the absence of an external magnetic field, the quasiparticle
density of states is given by

\bea g(E) = | x| Re \langle \frac{1}{\sqrt{x^2 - |f|^2}} \rangle,
\eea

where $x \equiv E/\Delta $, $f = \frac{3}{2} \left(1 -
\hat{k}_1^4 - \hat{k}_2^4 - \hat{k}_3^4\right) $ for the A-phase,
and $f = 1 - \hat{k}_3^4$ for the B-phase. Here $\langle ...
\rangle =(4 \pi )^-1 \int d \Omega ...$ denotes the angular average.
These quasiparticle densities of states are evaluated numerically
and shown in Fig. 2. In particular for the low-energy limit $|x|
\ll 1$ we find $g(E) \approx \pi |x|/4$ for the A-phase and $g(E)
\approx \pi |x|/8$ for the B-phase.

\begin{figure}[h]
\includegraphics[width=11.0cm]{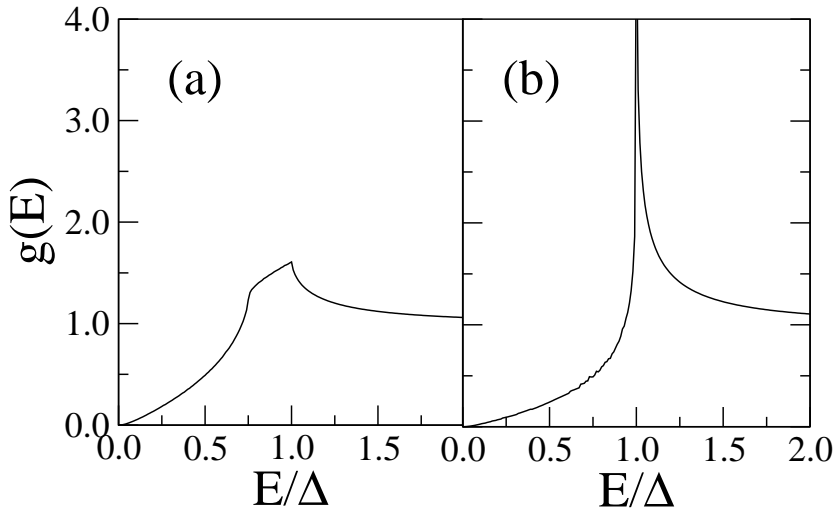}
\caption{
Quasiparticle density of states in a p+h-wave superconductor.
(a) A-phase, and (b) B-phase. 
}
\end{figure}

In the vortex state the quasiparticle density of states at $E = 0$
is given by $\pi \langle |\vec{v} \cdot \vec{q}| \rangle /(4
\Delta) $ (A-phase) and $\pi \langle |\vec{v} \cdot \vec{q}|
\rangle /(8 \Delta)$ (B-phase), where $\vec{v} \cdot \vec{q}$ is
the Doppler shift.\cite{volovik,won3} Note that 
in this case $\langle ... \rangle $
denotes the average over the unit cell of the vortex lattice and
over the nodal lines and points on the Fermi surface. For the
field dependence we find

\bea g_A(\vec{H}) & = & \frac{1}{2} \frac{v \sqrt{eH}}{\Delta} I_A
(\theta
, \phi ) \ \ \ \ \ \ \ \rm(A-phase), \\
g_B(\vec{H}) & = & \frac{1}{4} \frac{v \sqrt{eH}}{\Delta} I_B
(\theta , \phi ) \ \ \ \ \ \ \ \rm(B-phase), \eea with \bea
I_A(\theta , \phi ) & = & \sin \theta + ( 1 - \cos^2 \theta
\cos^2 \phi )^{1/2} + ( 1 - \cos^2 \theta \sin^2 \phi )^{1/2},\\
I_B (\theta , \phi ) & = & \sin \theta . \eea For the B-phase it
was assumed here that the point nodes are aligned parallel to
[001].

The corresponding specific heat, the spin susceptibility, and the
superfluid density in the clean limit and at ultra-low
temperatures are given by \cite{won4} \bea C_S / (\gamma_N T ) & =
& g (\vec{H}) , \\ \chi_s/\chi_N &=& g(\vec{H}), \\ \rho_s(H) /
\rho_S (0) &=& 1 - g(\vec{H}) \ \ \ \ \ \ \  \rm(A-phase), \\
\rho_s(H) /
\rho_S (0) &=& 1 - 3 g(\vec{H}) \ \ \ \ \  \rm(B-phase).\eea 
The superfluid density in the A-phase is isotropic and given by
Eq. 10. However, in the B-phase Eq. 11 is only valid if the supercurrent
flows parallel to
the z axis (parallel to the nodal
direction).
Therefore it will be of great interest to
study the Knight shift in the vortex state.

\section{Angle dependent magnetothermal conductivity}

In order to analyze the thermal conductivity it is necessary to
consider the effects of impurity scattering. Unlike for the s+g
order parameter, the effects of disorder in the p+h-wave
superconductor are more conventional.\cite{yuan,maki2} Here we
consider impurity scattering in the unitary limit.

Let us first focus on the self-consistent equation for impurity
scattering, given by $i C_0 \Delta = \lim_{\omega \rightarrow 0}
\tilde{\omega }$, where $\tilde{\omega}$ is the
disorder-renormalized frequency.\cite{won3} For the A-phase, this
leads to 
\bea 
C_0 = \frac{2 \Gamma}{\Delta} \left( C_0 \langle \ln
\left( \frac{2}{\sqrt{C_0^2 + x^2}} \right) \rangle + \langle x
\tan^{-1} \frac{x}{C_0} \rangle \right)^{-1} ,
\eea where 
$\Gamma$
is the quasiparticle scattering rate in the normal state, and $x
\equiv |\vec{v} \cdot \vec{q} | /\Delta $. For the B-phase the
prefactor 2 on the right-hand side of the above equation is
replaced by 4. In the superclean limit $\langle x \rangle \gg
C_0$ Eq. 12 leads to 
\bea C_{0A} &=& \frac{4 \Gamma}{\pi
\Delta} \langle x \rangle^{-1} \ \ \ \ \ \rm(A-phase),\\ 
C_{0B} &=&
\frac{8 \Gamma}{\pi \Delta} \langle x \rangle^{-1} \ \ \ \ \
\rm(B-phase)\eea 
On the other hand, in the clean limit we arrive at
\bea C_{0A}^2 \ln \frac{2}{C_{0A}} & = & \frac{2 \Gamma}{\Delta} -
\frac{\langle x^2 \rangle}{2},\ \ \ \ \ \ \rm(A-phase)\\ C_{0B}^2 \ln
\frac{2}{C_{0B}} & = & \frac{4 \Gamma}{\Delta} - \frac{\langle x^2
\rangle}{2},\ \ \ \ \ \ \rm(B-phase).\eea

In the absence of a magnetic field, the thermal conductivity
exhibits universal heat conduction\cite{lee,sun} with
$\kappa_{00}/T = (\pi^2 n)/(12 m \Delta ) $ in the A-phase. In the
B-phase there is universal heat conduction $\kappa_{00}/T = (\pi^2
n)/(8 m \Delta ) $ only when the point nodes are aligned
parallel to the
heat current $J_{\bf q}$. Here $\kappa_{00} $ is the thermal
conductivity in the limits $T \rightarrow 0$ and $\Gamma
\rightarrow 0$. Hence the experimental data of Ref. \cite{izawa2}
are consistent with p+h-wave superconductivity in $\rm
PrOs_4Sb_{12}$, but inconsistent with a s+g-wave order parameter.

In the vortex state, the thermal conductivity in the superclean
limit is given by 
\bea \frac{\kappa_{zz}}{\kappa_N} & = & \frac{
v^2 (eH)}{8 \Delta^2} \sin^2 \theta 
\ \ \ \ \rm(A-phase),\\
\frac{\kappa_{zz}}{\kappa_N} & = & \frac{3 v^2 (eH)}{64
\Delta^2} \sin^2 \theta 
\ \ \ \ \rm(B-phase),\eea
where $J_{\bf q} \parallel z$. 
Here the thermal conductivity in the normal state is given by
$\kappa_N = (\pi^2 n T)/(3 m \Gamma)$.
For the B-phase we have assumed that the nodes are aligned to [0 0 1]. 
When the nodes are along [1 0 0] or [0 1 0], $\kappa_{zz}$
is at least smaller by a factor of 10$\sim$50.
Note that $\kappa_{zz} \sim H \sin^2 \theta$ for both the A-phase and the 
B-phase. Therefore, the observed H-linear thermal conductivity 
$\kappa_{zz}$ at $T < 0.3K$ \cite{izawa2} follows from Eqs. 17 and 18. 
This implies that the crystals used in these measurements are in the 
superclean limit at sufficiently low temperatures. 

\begin{figure}[h]
\includegraphics[width=9.0cm]{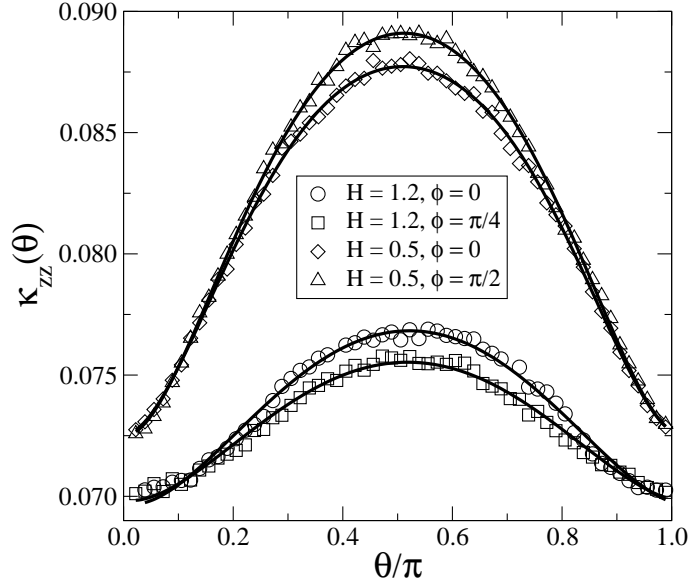}
\caption{
Angle dependence of the thermal conductivity in an applied magnetic field. 
The symbols are experimental data from Ref. \cite{izawa2}, and the solid
lines are fits to $ H F(\theta )$, as discussed in the text. 
}
\end{figure}

In the clean limit the expressions for the thermal conductivity become
\bea
\frac{\kappa_{zz}}{\kappa_{00}} & = & 1 + \frac{3 v^2 (eH)}{40 \Gamma \Delta} 
\ln \left( \sqrt{\frac{2\Delta}{\Gamma}} \right) \sin^2 \theta 
\ln \left( \frac{\Delta}{v\sqrt{eH} \sin \theta
} \right) \ \ \ \ \ \rm(A-phase),\\
\frac{\kappa_{zz}}{\kappa_{00}} & = & 1 + \frac{v^2 (eH)}{12 \Gamma \Delta}
\ln \left( \sqrt{\frac{2\Delta}{\Gamma}} \right) \sin^2 \theta
\ln \left( \frac{\Delta}{v\sqrt{eH} \sin \theta 
} \right) \ \ \ \ \ \rm(B-phase),
\eea
where $\kappa_{00}$ is the thermal conductivity in the limit of universal 
heat conduction.
Hence, in the clean limit the field-dependent part of $\kappa_{zz}$ is
given by $\kappa_{zz} \sim H F(\theta )$, where $F(\theta ) = 
\sin^2 \theta \ln (C/ \sin \theta )$ and $C = \Delta /v\sqrt{e H}$.
From these equations we observe that the $\theta$-dependence 
of the leading terms is the same for the A-phase and B-phase. 
In Fig. 3, we plot the measured $\theta$-dependence of $\kappa_{zz}$
in $\rm PrOs_4Sb_{12}$ \cite{izawa2} at various applied magnetic fields and 
azimuthal angles $\phi$. Fits of this data to $H F(\theta )$ are 
shown as solid lines, suggesting that these samples are in the clean limit.
In these fits $C = 5$
and $C = 3$ are obtained
for $H = 0.5 T$ and $H =  1.2 T$ respectively. Thus, using 
the weak-coupling theory gaps $\Delta_A = 4.2 K$ and $\Delta_B = 3.5 K$
for the A- and B-phase,
we deduce $ v=0.96 \times 10^7 cm/sec$ and $\Gamma \simeq 0.1 K$.
These values are reasonable,\cite{footnote} indicating that the quasiclassical
approximation is reliable.

\section{Concluding Remarks} 

In this work, we have proposed a spin-triplet p+h-wave order parameter
to account for observed features in the superconducting phases of
$\rm PrOs_4Sb_{12}$. This model describes well the 
angle dependent thermal conductivity data by Izawa et al., 
Refs. \cite{izawa1} and \cite{izawa2}.
In order to be fully consistent,
we have discovered that the nodal directions of $\Delta ({\bf \rm k})$ in 
the B-phase have to be aligned parallel to the external magnetic field in 
the field-cooled configuration. This triplet superconductivity in 
$\rm PrOs_4Sb_{12}$ is not surprising since many other heavy-fermion 
superconductors
appear to have spin-triplet order parameters, including 
$\rm UPt_3$, $\rm U Be_{13}$, $\rm URu_2Si_2$, and $\rm UNi_2Al_3$.
\cite{tou2} The interesting
dependence of the nodal points in $\Delta ({\bf \rm k})$
on the external magnetic field deserves further study.

We are grateful to H. Tou for useful correspondence about his NMR
measurements on $\rm PrOs_4Sb_{12}$, and to P. Thalmeier for useful
conversations. K.M. is grateful for the hospitality of the
Max-Planck Institute for the
Physics of Complex Systems at Dresden, where part of this work was performed.
S.H. acknowledges financial 
support by the NSF under Grant No. 
DMR-0089882.

\end{document}